# Clusters of Cosmic Rays above 4.10$^{19}$ eV

A.A. Mikhailov, N.N. Efremov.
*Yu.G. Shafer Institute of Cosmophysical Research and Aeronomy, 31 Lenin Ave., 677980 Yakutsk, Russia*
Presenter: A. Mikhailov (mikhailov@ikfia.ysn.ru), rus-mikhailov-AA-abs2-HE14-poster

Arrival directions of cosmic rays with the energy E>4.10$^{19}$ eV are analyzed on the basis of the Yakutsk and AGASA extensive air shower arrays. It is supposed that the clusters can be formed as a result of fragmentation of superheavy nuclei. The consequences of this supposition compare with experimental data. Part of clusters with E>2.10$^{19}$ eV correlate with the nearest pulsars. The new effect is found – clusters intensify a feature of cosmic rays distribution.

## 1. Introduction

By using AGASA array data in the region of ultrahigh energy E>4.10$^{19}$ eV, 7 clusters have been found [1]: 1 triplet and 6 doublets. The distance between showers in the clusters is <2.5°. At that energies by Yakutsk EAS array data 2 clusters-doublets are also detected [2]. The distance between showers in the clusters is <5°. In [1,3] it is suggested that the clusters are formed from the neutral particles, in [4] they are as a result of focusing of charged particles by the magnetic field, in [5] as a stochastic process. Previously we showed [6] that E>4.10$^{19}$ eV cosmic rays were most likely superheavy nuclei. The clusters can be formed as a result of decay (spontaneous or as a result of interactions with the gas) into the more light nuclei [2]. It explains two facts observed: 1) the isotropic distribution of clusters, 2) the correlation of particle in the clusters. According to this hypothesis, among the secondary particles formed, the particle of the highest energy must arrival first. Consider this supposition.

## 2. Discussion

We consider Yakutsk EAS array data beginning with the energy 10$^{18.7}$ eV and AGASA array data with E>4.10$^{19}$ eV [1]. 1863 EAS' with zenith angles < 60° for the period 1974 to 2002 by Yakutsk data have been analyzed. Among them we found 5157 doublets (clusters with the large number of particles are not considered). The distance between EAS' in the doublets is <5°. Let the doublet particles have the energies E$_1$ and E$_2$ and the time in the path from the formation position to the Earth t$_1$ and t$_2$. We separated the total number of doublets into 2 parts: the doublets N$_1$(E$_1$) whose particle of E$_1$>E$_2$ arrives first to the Earth (t$_1$<t$_2$), N$_2$(E$_1$) are doublets whose particle of E$_1$>E$_2$ arrives second (t$_1$>t$_2$). The shower energy in experimental data is determined with some error, therefore it is difficult to determine what of doublet particles really has the higher energy. In this connection, the doublets having particles with energies close to one another within the limits of lg(E$_1$/E$_2$)<0.2 have been excluded from consideration. Under consideration of AGASA data according to the above criterion, the doublet A6 [1] was excluded.

Fig.1 presents a ratio of the number of these doublets (circles are Yakutsk, squares are AGASA) R= N$_1$(E$_1$)/N$_2$(E$_1$), where N$_1$(E$_1$) are doublets whose particle with E$_1$>E$_2$ arrives first to the Earth (t$_1$<t$_2$); N$_2$(E$_1$) are doublets whose particle with E$_1$>E$_2$ arrives second (t$_1$>t$_2$). As can be seen, at E<2.10$^{19}$ эB the doublets independent of leading particle energy arrive uniformly, on the average. At E>2.10$^{19}$ eV the number of doublets, whose particle with E$_1$>E$_2$ arrives first, (t$_1$<t$_2$) is ~ 70% for the Yakutsk array, and ~ 80% for AGASA.

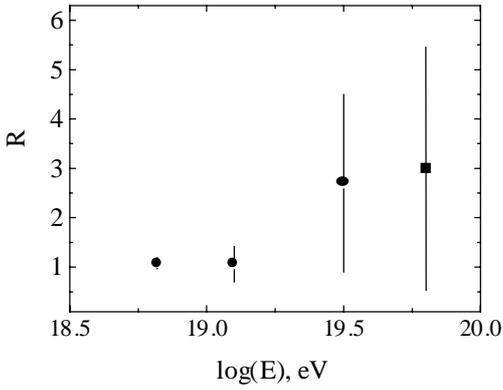

Fig.1. Ratio of the number of doublets $R=N_1(E_1)/N_2(E_1)$. $N_1(E_1)$ are doublets whose particle with $E_1>E_2$ arrives first ($t_1<t_2$); $N_2(E_1)$ are doublets whose particle with $E_1>E_2$ arrives second ($t_1>t_2$). ● - Yakutsk, ■ - AGASA.

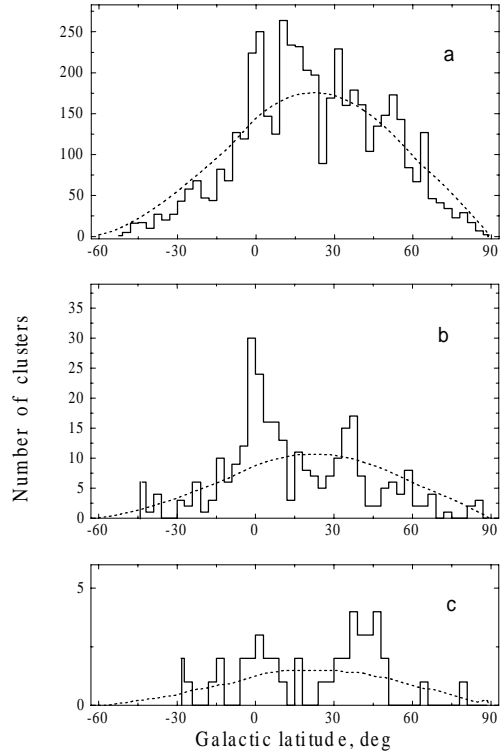

Fig.2. The doublet distribution in galaxy latitude for energy intervals: a) $10^{18.7} - 10^{19}$ eV, b) $10^{19} - 10^{19.3}$ eV, c) $>10^{19.3}$ eV.

Thus, in spite of large statistical errors in the determination of R the conclusion can be made, that the doublets at $E>2.10^{19}$ eV are most probably formed as a result of superheavy nuclei decay. Probably, the doublets at $E<2.10^{19}$ эВ are formed by other means than at $E>2.10^{19}$ eV.

Next, we will attempt to clarify origin doublets at $E<2.10^{19}$ eV. We consider the distribution of doublets in galactic latitude for 3 energy intervals: a) $10^{18.7} - 10^{19}$ eV – 4822 doublets (1387 EAS'), b) $10^{19} - 10^{19.3}$ eV – 293 doublets (260 EAS'), c) $>10^{19.3}$ eV -42 doublets (59 EAS').

Fig.2 demonstrates the number of doublets in $\Delta b=3°$ latitude intervals (as coordinates of doublets we take latitudes averaged for the doublets particles). In the energy range $E=10^{18.7} - 10^{19}$ eV (Fig.2a) at $|b|<3°$ the excess of the number of doublets observed over the expected one in the case of isotropy is $11\sigma=(474-288.9)/\sqrt{288.9}$. The number of doublets expected is found by the Monte-Carlo method taking into account the array exposure at the celestial sphere. At latitudes of $9°<b<24°$ a maximum in the doublet distribution is also observed. The detailed analysis shows that this maximum is confined in galactic longitudes within $120°<l<150°$. This maximum in the doublet distribution repeats the maximum in the shower distribution at E $\sim 10^{19}$ eV in the above coordinates [7].

From Fig.2b it can be seen that at $E=10^{19} - 10^{19.3}$ eV and $|b|<3°$ the number of doublets exceeds the number of doublets expected by $8.7\sigma=(54-17.5)/\sqrt{17.5}$. At these energies the distribution of EAS' in galactic latitude b is presented in Fig.3a. In the EAS' distribution several maxima are observed: at $|b|<3°$ (the excess is $2.7\sigma$), at $15°<b<21°$ and etc, but in the doublet distribution (Fig.2b) the maximum is seen at $|b|<3°$, only. It can mean that the maximum number of doublets appear where there is the particle flux extended in latitude b. Wherever maxima in the particle distribution are narrow, random (Fig.3a), maxima in the doublet distribution are not formed.

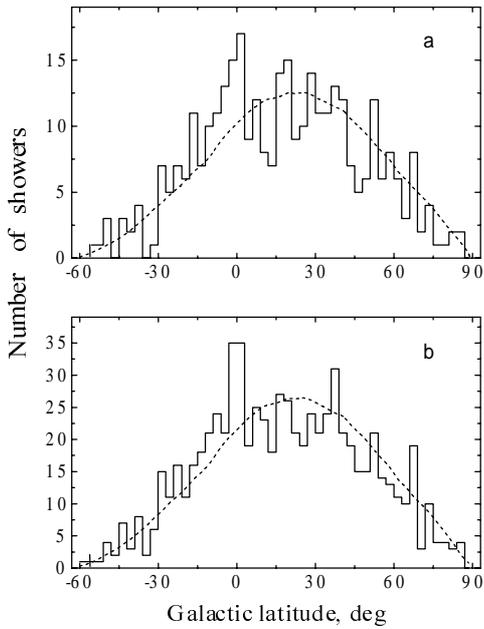

Fig.3. The EAS' distribution in galaxy latitudes: a) $10^{19} - 10^{19.3}$ eV, b) $10^{18.7} - 10^{19.6}$ eV.

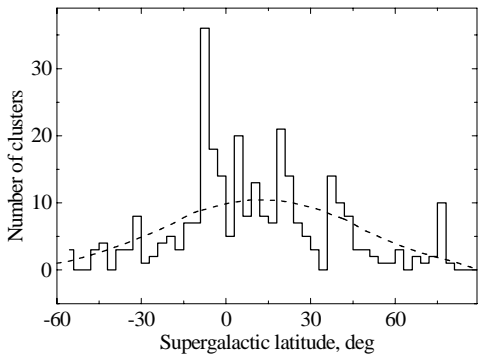

Fig.5. The doublet distribution in Supergalaxy latitude at $10^{19} - 10^{19.3}$ eV.

origin.

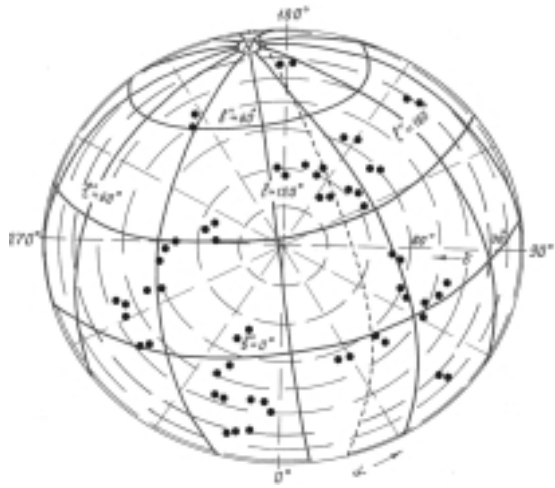

Fig.4. Distribution of doublets with E>$10^{19.3}$ eV on celestial sphere: $\delta$, $\alpha$ - declination and right ascension, b, l – galactic latitude and longitude. Dashes curve – Supergalaxy plane.

As can be seen in Fig.2a,b the maximum in doublets distribution at E ~ $10^{19}$ eV points to the galactic plane. In this connection note that earlier we had found in the energy range E= $10^{18.9} - 10^{19.6}$ eV the excess particle flux from the galactic plane [8,9]. This result is verified by new data, at |b|<3° the excess of the number of EAS' observed over the expected one in the case of isotropy is $4.1\sigma = (70\text{-}43)/\sqrt{43}$ (Fig.3b).

Fig. 2c shows that at E>$10^{19.3}$ eV the doublet distribution is more isotropic, but at b ~ 0° there is maximum in EAS' distribution. In Fig.4 the distribution of these 42 doublets is shown on the equal-exposure map of celestial sphere. 31 doublets from 42 are <9° (at <6° - 23 doublets) from pulsars [10]. Chance probability is ~ 0.2. All pulsars are situated <3 kpc from Earth. The distribution of doublets is limits by galactic latitudes in the main -30°<b<50°. From here we can conclude that most likely doublets with E>$10^{19.3}$ eV has galactic origin.

So, the distribution of doublets (clusters) repeats the distribution of showers, at the same time intensifying of their maximum by several times. Therefore, arrival directions of doublets can point to the cosmic ray anisotropy. Also the distribution of doublets points their galactic origin.

Fig.5 presents the doublets distribution at E=$10^{19}$ – $10^{19.3}$ eV in Supergalaxy latitude $b_{SG}$. Their maxima do not coincide with Supergalaxy plane (see also Fig.4), whence it follows that the Supergalaxy plane is no source of such energy particles. Early we showed that at E ~ $10^{19}$ eV maximum of EAS' distribution it is observed near Supergalaxy plane, but this maximum was limited by galactic latitudes -30°<b<40° [7]. From here we concluded that this maximum of EAS' has a galactic origin. Thus, the doublet distribution in Supergalaxy coordinates confirms our old result [7].

## 3. Conclusions

At E>2.$10^{19}$ eV clusters are most likely formed as a result of the superheavy nuclei decay in our Galaxy. The clusters intensify a feature of cosmic rays distribution and can point to their maxims.

## 4. Acknowledgements

The work has been supported by RFBR (project N 04-02-16287). The Yakutsk EAS array was supported by the Ministry of Training of the Russian Federation, project no.01-30.